\documentclass{PoS}
\usepackage{graphicx}
\usepackage{amsmath}
\usepackage{amsfonts}
\usepackage{amssymb}

\newcommand{\ie}{{\it i.e.}}

\newcommand{\eg}{{\it e.g.}}

\newcommand{\fig}{Fig.}

\newcommand{\Ref}{Ref.}

\newcommand{\figu}[1]{\fig~\ref{fig:#1}}

\title{Fundamental physics with high-energy cosmic neutrinos today and in the future}

\ShortTitle{High-Energy Fundamental Neutrino Physics}

\author{Carlos A. Arg\"uelles \\
       Massachusetts Institute of Technology\\
        E-mail: \email{caad@mit.edu}}

\author{\speaker{Mauricio Bustamante}\\
        Niels Bohr International Academy \& DARK, Niels Bohr Institute, University of Copenhagen\\
        E-mail: \email{mbustamante@nbi.ku.dk}}

\author{Ali Kheirandish \\
       Department of Physics \& Wisconsin IceCube Particle Astrophysics Center, \\ University of Wisconsin - Madison\\
        E-mail: \email{akheirandish@icecube.wisc.edu}}
       
\author{Sergio Palomares-Ruiz \\
       Instituto de F\'isica Corpuscular, CSIC-Universitat de Val\`encia\\
        E-mail: \email{sergiopr@ific.uv.es}}

\author{Jordi Salvado \\
       Department de F\'isica Qu\`{a}ntica i Astrof\'isica and Institut de Ci\`{e}ncies del Cosmos, \\ Universitat de Barcelona\\
        E-mail: \email{jsalvado@icc.ub.edu}}
       
\author{Aaron C.~Vincent \\
       Arthur B.~McDonald Canadian Astroparticle Physics Research Institute, Department of Physics, Engineering Physics and Astronomy, Queen's University,\\
       {\rm and} Perimeter Institute for Theoretical Physics \\
        E-mail: \email{aaron.vincent@queensu.ca}}
       
\abstract{The astrophysical neutrinos discovered by IceCube have the highest detected neutrino energies --- from TeV to PeV --- and likely travel the longest distances --- up to a few Gpc, the size of the observable Universe.  These features make them naturally attractive probes of fundamental particle-physics properties, possibly tiny in size, at energy scales unreachable by any other means.  The decades before the IceCube discovery saw many proposals of particle-physics studies in this direction.  Today, those proposals have become a reality, in spite of astrophysical unknowns.  We will showcase examples of doing fundamental neutrino physics at these scales, including some of the most stringent tests of physics beyond the Standard Model.  In the future, larger neutrino energies --- up to tens of EeV --- could be observed with larger detectors and further our reach.}

\FullConference{36th International Cosmic Ray Conference -ICRC2019-\\
		July 24th - August 1st, 2019\\
		Madison, WI, U.S.A.}

\begin{document}


\section{Introduction}

What is Nature like at its most fundamental level?  What are its building blocks and how do they interact?  What are its organizing principles?  During the last century, we steadily found deeper answers using increasingly powerful particle accelerators that revealed fundamental particles, interactions, and  symmetries.  Yet, ample territory remains unexplored at higher energies, ripe for discoveries.  Today, accelerators still churn out valuable data, but, so far, fail to guide us in furthering our view of fundamental physics.  Observing particle processes at higher energies would provide guidance, but they lie beyond the reach of current accelerator technology. 

Fortunately, Nature itself provides a way forward: we can turn from human-made particle accelerators to naturally occurring cosmic accelerators.  These are extreme astrophysical phenomena --- \eg, compact-object mergers, supernovae, black holes and their jets --- that emit particles with energies millions of times higher than accelerators.  

High-energy cosmic neutrinos, discovered by the IceCube Neutrino Observatory~\cite{Gaisser:1994yf, Aartsen:2013jdh}, are particularly incisive probes of particle physics~\cite{Ahlers:2018mkf, Ackermann:2019cxh}.  They reach Earth, potentially after crossing the size of the observable Universe, bringing information on high-energy processes.  Because neutrinos barely interact with matter on their way, this information reaches us largely unperturbed.  

Here, we take stock of the current and near-future landscape of testing particle physics with high-energy cosmic neutrinos.  Our aim is to draw attention to the power of these tests, accessible already today, by showing a bird's eye view of the landscape.  In our presentation, we favor breadth over depth; details can be found in, \eg, \Ref\ \cite{Ahlers:2018mkf}.  To help guide efforts, we introduce a classification scheme of models of new neutrino physics, \ie, of physics beyond the Standard Model.


\section{Why use high-energy cosmic neutrinos?}

\begin{figure}[t!]
 \centering
 \includegraphics[width=1.0\textwidth]{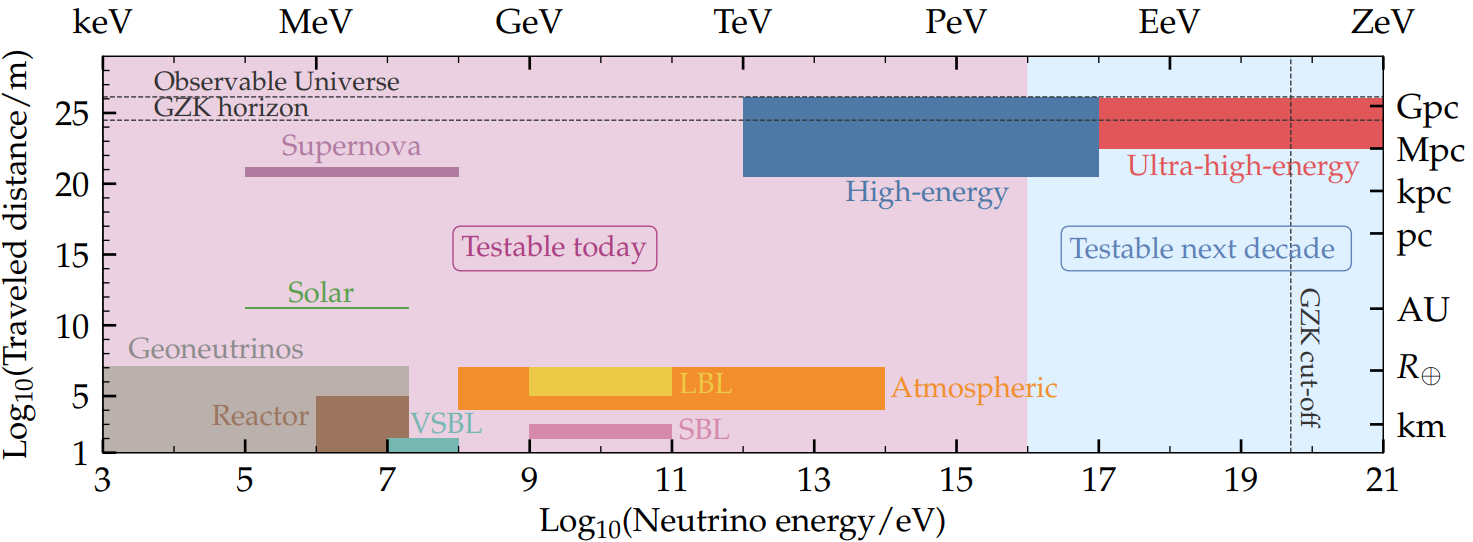}
 \caption{Energy and distance scales of neutrinos from different sources.  For comparison, GZK (Greisen-Zatsepin-Kuzmin) refers to the maximum distance and energy that ultra-high-energy cosmic rays can reach.  Figure adapted from \Ref\ \cite{Ackermann:2019cxh}.}
 \label{fig:scales}
\end{figure}

Figure\ \ref{fig:scales} shows the energy and distance scales of neutrinos from different sources.  We focus on high-energy and ultra-high-energy neutrinos, both of cosmic origin, which reach the highest values on both scales.  They can probe numerous physics effects.  There are at least four reasons why high-energy cosmic neutrinos are well-suited to test particle physics:
\begin{itemize}
 \item
  {\bf They have the highest neutrino energies detected.}  Because cosmic neutrinos reach TeV--PeV energies, they can probe neutrino physics that is suppressed by a high-energy scale, previously inaccessible.  Conveniently, the intensity of many new-physics models is expected to grow with neutrino energy.  For comparison, the most energetic neutrino made in a particle accelerator reached about 350~GeV\ \cite{Tzanov:2005kr}.  High-energy cosmic neutrinos can probe physics at scales of $\sqrt{s} \sim 100$~TeV, comparable to the envisioned Future Circular Collider (see \figu{cross_section}).
 \item
  {\bf They travel the longest distances.}  The bulk of high-energy cosmic neutrinos is likely made in extragalactic sources.   Their ability to traverse the size of the observable Universe en route to Earth allows for tiny new-physics effects, otherwise unobservable, to accumulate during the long trip and become large enough to be detectable upon reaching Earth.
 \item
  {\bf They are weakly interacting.}  Because neutrinos have tiny cross sections, after being produced they are not expected to interact while propagating to Earth, until they are detected.  However, new-physics effects could change this.  The absence of standard interactions during propagation removes one layer of uncertainty when looking for new physics.
  \item
   {\bf They have a quantum number not shared by cosmic rays or photons: flavor.} Different neutrino production mechanisms at the sources result in different numbers of neutrinos of each flavor.  Thus, flavor can be used as a discriminant. Standard neutrino mixing during propagation constrains the possible flavor combinations at Earth to a small region.  However, new physics effects could significantly change this picture, provided detection systematics on flavor identification are under control\ \cite{Mena:2014sja, Palomares-Ruiz:2015mka, Li:2016kra}. 
\end{itemize}

Broadly stated, the size of the effect introduced by many new-physics models grows as $\sim \kappa_n E_\nu^n L$, where $\kappa_n$ is a model-specific coupling, $E_\nu$ is the energy of the neutrino, and $L$ is the distance traveled by the neutrino.  The constant $n$ is model-specific, \eg, in models where neutrinos decay\ \cite{Gelmini:1982rr, Baerwald:2012kc}, $n=-1$, while in models with violation of CPT or Lorentz invariance\ \cite{Colladay:1998fq, Kostelecky:2003cr}, $n \geq 0$.  Using neutrinos with $E_\nu \sim$~PeV and $L \sim$~Gpc, we are sensitive to tiny couplings, \ie, $\kappa_n \sim 4 \cdot 10^{-47} (E_\nu/{\rm PeV})^{-n} (L/{\rm Gpc})^{-1}$~PeV$^{1-n}$, orders of magnitude smaller than the limits on the couplings obtained using atmospheric neutrinos\ \cite{Aartsen:2017ibm}.


\section{High-energy cosmic neutrinos: a quick review}

IceCube has firmly detected a flux of TeV--PeV neutrinos of cosmic origin\ \cite{Aartsen:2013jdh, Aartsen:2014gkd}.  IceCube monitors 1~km$^3$ of Antarctic ice using photomultipliers, in search for the dim flashes of Cherenkov light that are produced when high-energy neutrinos collide with a nucleon of the ice.  From the spatial and temporal profiles of the light signals, IceCube reconstructs the deposited energy, arrival direction and time, and flavor; we comment on the associated measurement uncertainties later.

High-energy neutrino sources are believed to be made in hadronic accelerators.  Candidates include blazars, gamma-ray bursts, superluminous supernovae, tidal disruption events, millisecond pulsars, and starburst galaxies (see, \eg, \Ref\ \cite{Anchordoqui:2013dnh}).  In them, protons and other nuclei are accelerated up to energies of $E_p \sim 10^{12}$~GeV; their spectrum is expected to be $\propto E_p^{-2}$. 

In the sources, protons with energies of tens of PeV interact with ambient matter and photons to produce pions which subsequently decay to neutrinos with TeV--PeV energies.  For instance, in proton-photon interactions, the most likely production channel is $p + \gamma \to \Delta^+(1232) \to \pi^+ + n$, followed by $\pi^+ \to \mu^+ + \nu_\mu$ and $\mu^+ \to \bar{\nu}_\mu + e^+ + \nu_e$.  Each neutrino carries about 5\% of the parent cosmic-ray energy.  Neutrinos thus produced inherit a power-law energy spectrum from the parent protons, \ie, $\propto E_\nu^{-\alpha}$.  The value of $\alpha$ and the flux level depend on the source details, on the target photon spectrum, and on the mass composition of nuclei in the source; see, \eg, \Ref\ \cite{Biehl:2017zlw}. 

High-energy cosmic neutrinos are predominantly extragalactic, since the distribution of their arrival directions is compatible with an isotropic distribution\ \cite{Aartsen:2018ywr}. To date, the search for their origin has revealed strong evidence for neutrino emission from the direction of a gamma-ray blazar\ \cite{IceCube:2018dnn, IceCube:2018cha}.  However, the origin of the majority of the cosmic neutrino flux remains unidentified.


\section{A bird's eye view of the landscape of new neutrino physics}
\label{section:landscape}

\begin{figure}[t!]
  \centering
  \includegraphics[width=0.45\linewidth]{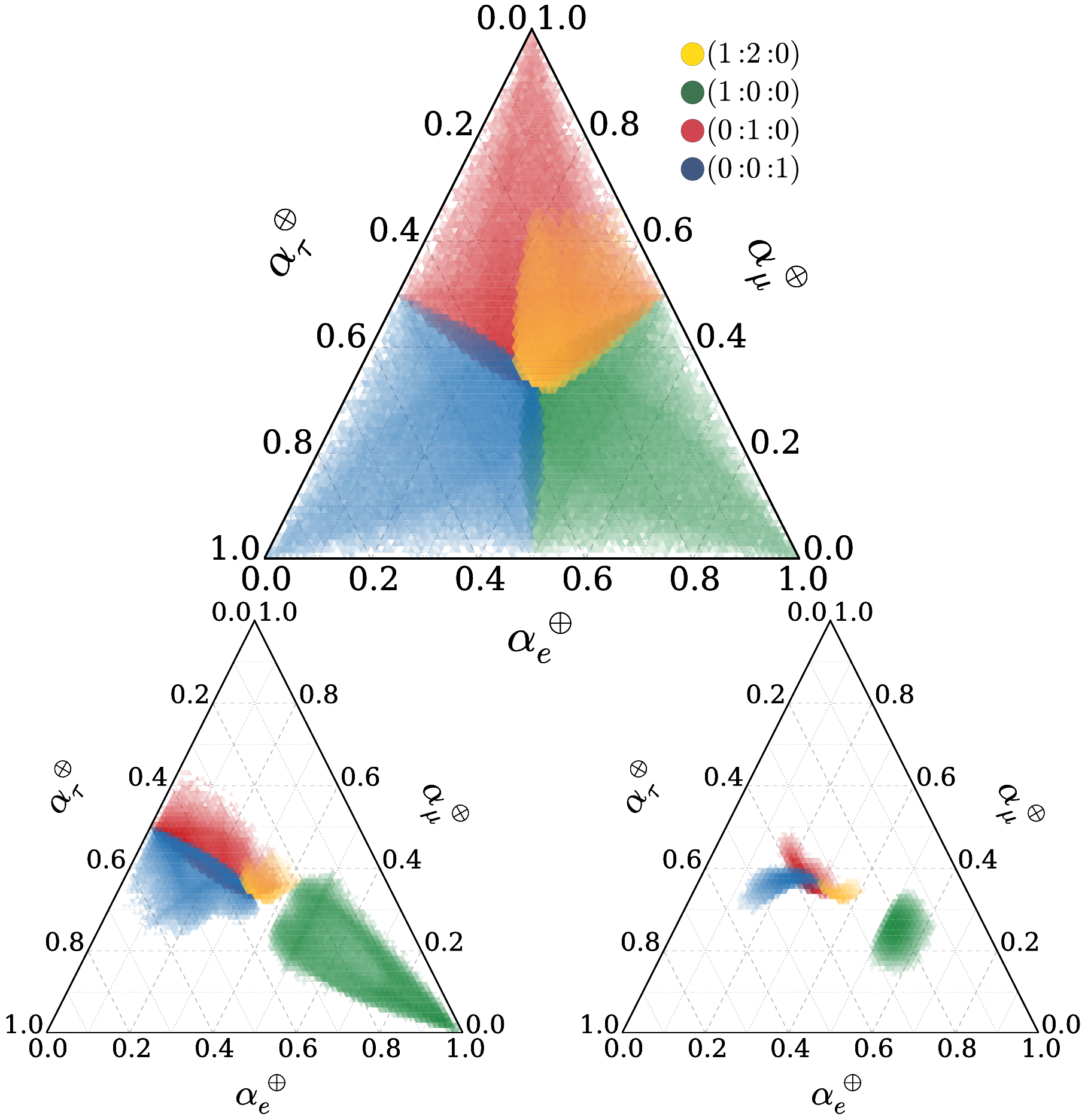}
  \includegraphics[width=0.45\linewidth]{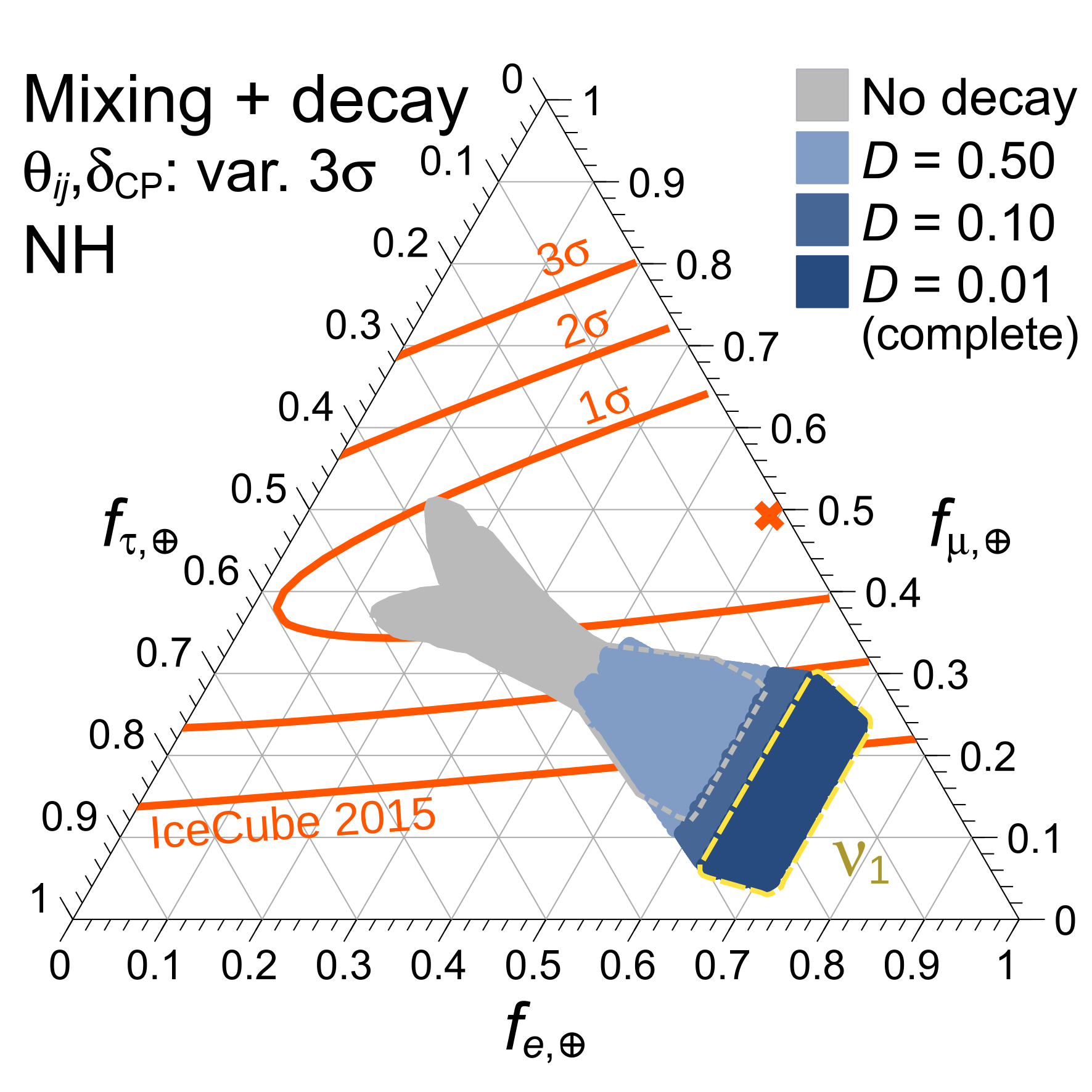}
  \caption{\label{fig:ternary_plot}Flavor composition of high-energy cosmic neutrinos at Earth. {\it Left:} Regions expected from different energy scales of Lorentz-invariance violation, starting from different flavor compositions at the sources\ \cite{Arguelles:2015dca}.  {\it Right:} Regions expected from neutrino decay, for any composition at the sources, with $D$ the fraction of non-decaying neutrinos that survive until reaching Earth\ \cite{Bustamante:2016ciw}.}
\end{figure}

\begin{figure}[t!]
 \begin{minipage}[t]{0.492\textwidth}  
  \centering
  \begin{minipage}[c][8cm][c]{\linewidth}
  \includegraphics[width=\linewidth]{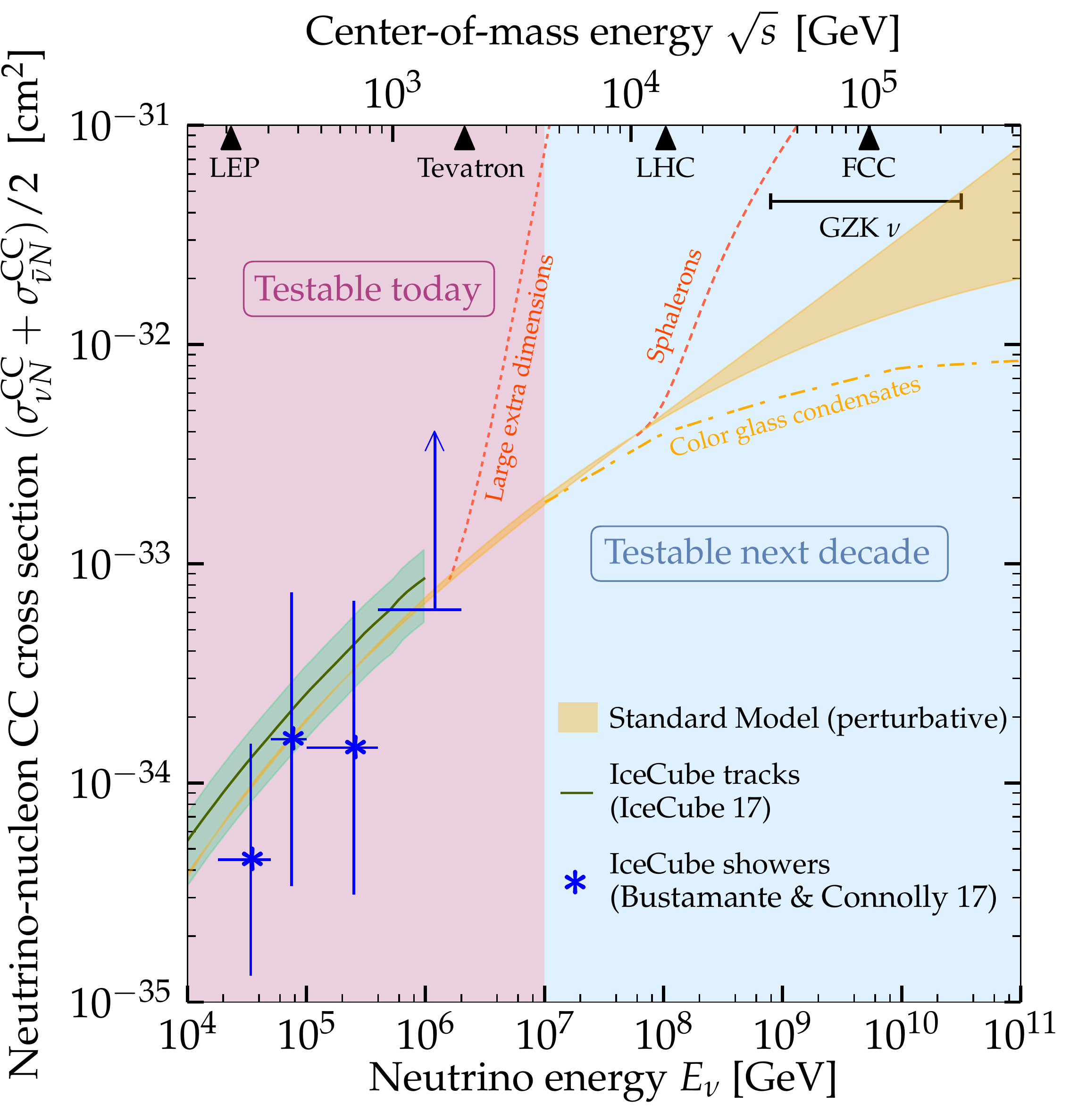}
  \end{minipage}
  \caption{\label{fig:cross_section}Measurements and predictions of the high-energy neutrino-nucleon cross section using IceCube data\ \cite{Bustamante:2017xuy, Aartsen:2017kpd}; see also \Ref\ \cite{Aartsen:2018vez}.  Updated versions of the color glass condensate model\ \cite{Arguelles:2015wba} still show significant differences.  Figure extracted from \Ref\ \cite{Ackermann:2019cxh}.}
\end{minipage}
\hspace*{0.1cm} 
\begin{minipage}[t]{0.492\textwidth}  
  \centering
  \begin{minipage}[c][8cm][c]{\linewidth}
  \includegraphics[width=\linewidth, clip=true, trim=1.0cm 0 1.2cm 0.7cm]{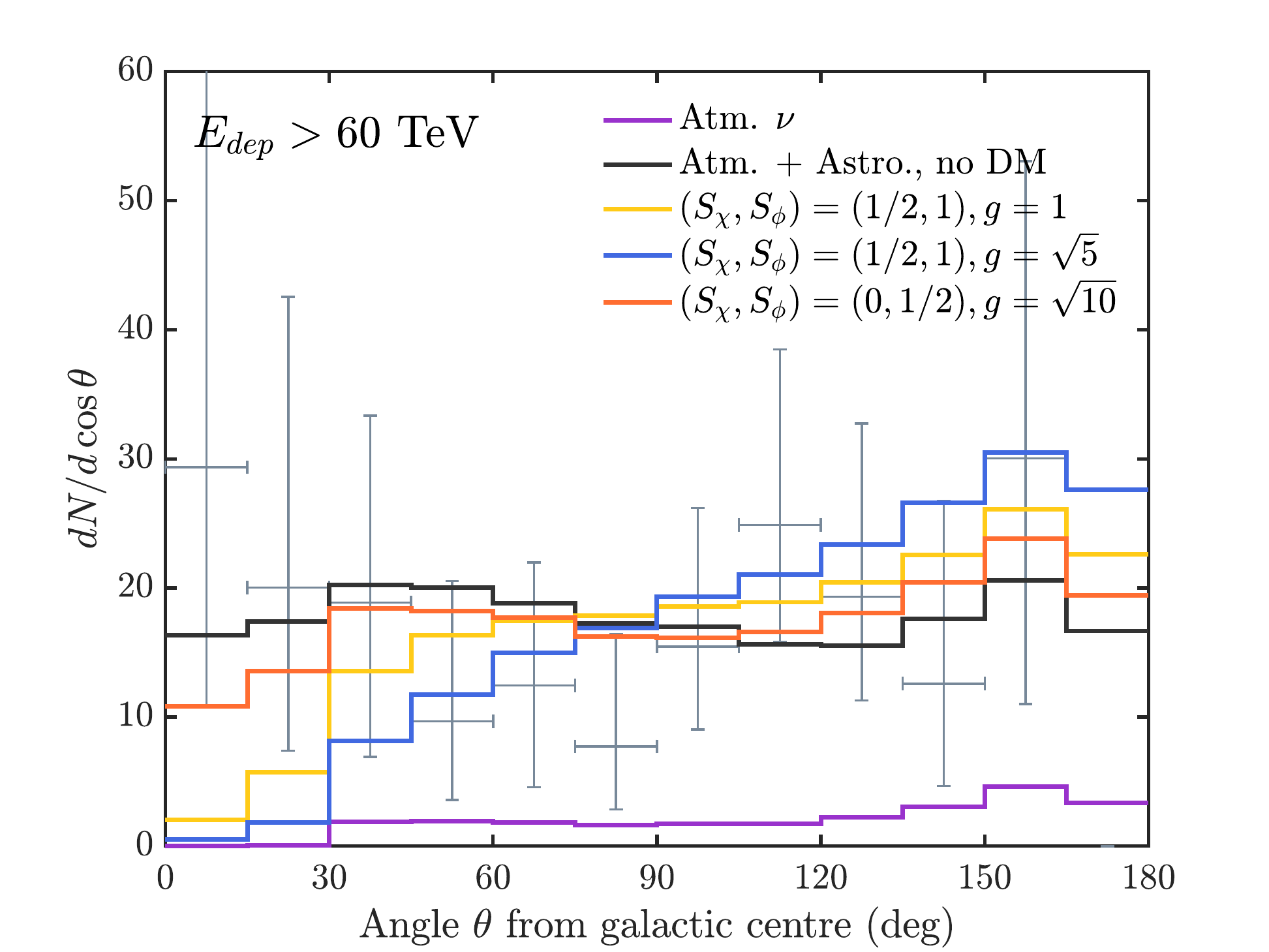}
  \end{minipage}
  \caption{\label{fig:nu_dm}Distribution of IceCube HESE events in the presence of neutrino interactions with different models of dark matter, as a function of angular distance from the Galactic Center.  Figure extracted from \Ref\ \cite{Arguelles:2017atb}.}
 \end{minipage}
\end{figure}

Some of the most important, open particle-physics questions that can be answered with high-energy cosmic neutrinos are: How do neutrino cross sections behave at high energies?  (See \figu{cross_section}.)  How do flavors mix at high energies?  What are the fundamental symmetries of Nature?  Are neutrinos stable?  What is dark matter?  Are there sterile neutrinos?  Are there hidden neutrino interactions with cosmic backgrounds?  In the following, we refer briefly to some of these questions; for details, see, \eg, \Ref\ \cite{Ackermann:2019cxh}.  To answer these questions, we focus on four prominent high-energy neutrino observables, measured at Earth by neutrino telescopes:

\begin{itemize}
 \item 
  {\bf Energy:} The standard expectation for the diffuse neutrino energy spectrum is a featureless power law.  However, new-physics effects may introduce unique spectral features, notably, dips and pile-ups, due, \eg, to new interactions between high-energy neutrinos and the relic background of low-energy neutrinos\ \cite{Ng:2014pca}, interactions between neutrinos and dark matter\ \cite{Arguelles:2017atb}, or leptoquarks affecting neutrino-nucleon interactions\ \cite{Dey:2017ede}.
 \item
  {\bf Arrival direction:} If the sources of cosmic neutrinos are numerous and dim, then the arrival directions of neutrinos at Earth should be isotropic.  Therefore, anisotropies in the neutrino sky could be indicative of new sources.  In the direction of large concentrations of matter --- \eg, the Galactic Center\ \cite{Arguelles:2017atb} (see \figu{nu_dm}) or satellite dwarf galaxies\ \cite{Cherry:2014xra} --- they could be indicative of new neutrino interactions with ordinary matter or with dark matter\ \cite{Arguelles:2017atb}.
 \item
  {\bf Flavor composition:}  We expect neutrinos to be produced with approximate flavor ratios of $(\nu_e:\nu_\mu:\nu_\tau)_{\rm S} = (1:2:0)_{\rm S}$.  En route to Earth, standard neutrino oscillations transform this to $(1:1:1)_\oplus$.  Figure\ \ref{fig:ternary_plot} shows that, even if the flavor ratios at the sources are different from $(1:2:0)_{\rm S}$, the flavor ratios at Earth can only occupy a small region of parameter space\ \cite{Bustamante:2015waa}.  For instance, neutrino decay\ \cite{Baerwald:2012kc, Bustamante:2016ciw}, Lorentz-invariance violation\ \cite{Arguelles:2015dca, Ahlers:2018yom}, and interactions with dark matter\ \cite{Farzan:2018pnk} could introduce large deviations in the flavor ratios.
 \item
  {\bf Timing:}  If transient astrophysical events --- \eg, blazar flares, gamma-ray bursts, tidal disruption events --- make high-energy neutrinos, they should arrive at Earth at the same time as the electromagnetic signals from these events, barring small differences in their production times.  However, Lorentz-invariance violation\ \cite{Kostelecky:2003cr} or new interactions between high-energy neutrinos and the relic neutrino background\ \cite{Murase:2019xqi} could introduce significant delays in the arrival times between neutrinos and photons, and between neutrinos of different energies.
\end{itemize}

\begin{figure}[t!]
 \centering
 \includegraphics[width=0.7\textwidth]{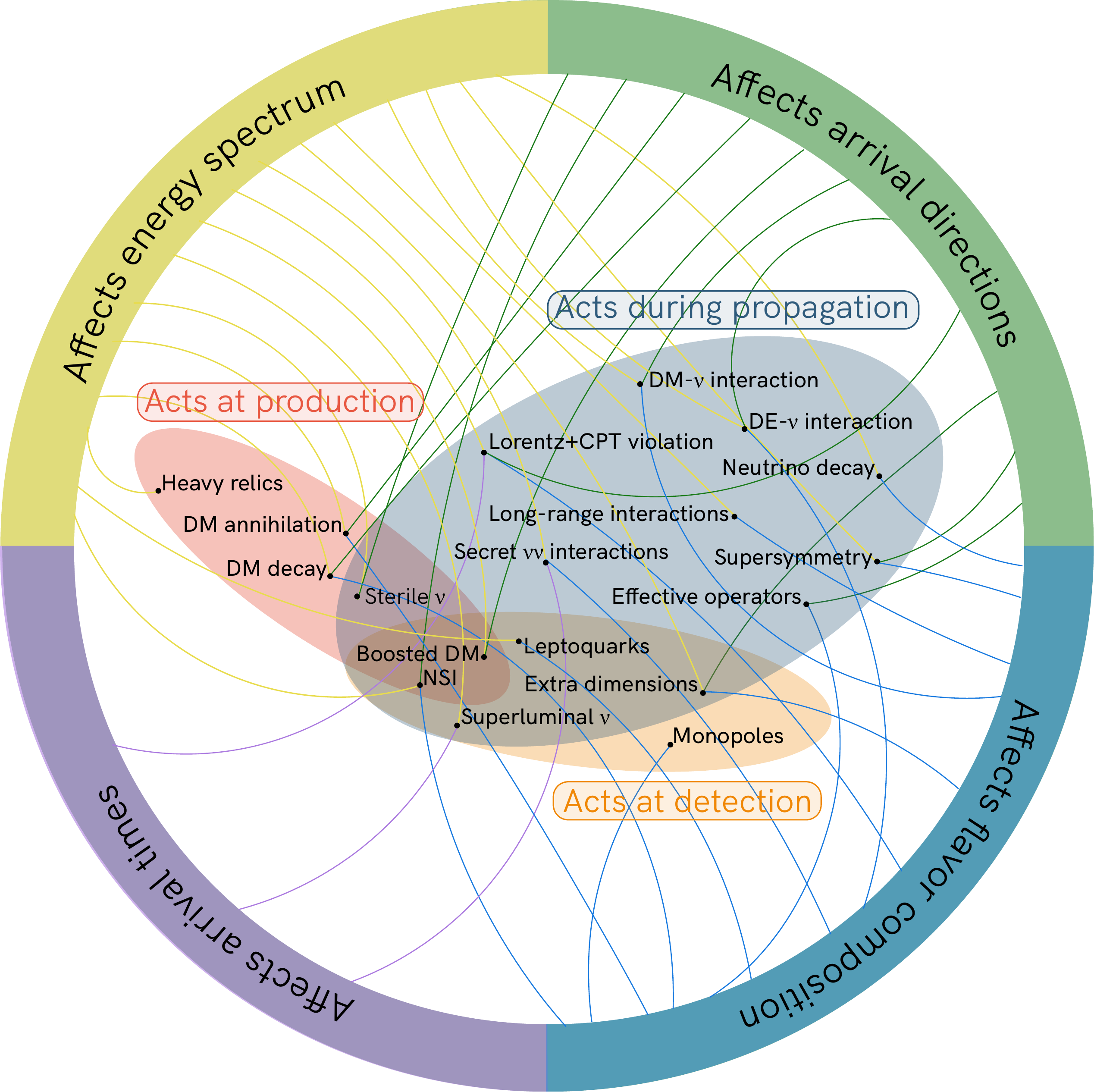}
 \caption{Classification of models of new neutrino physics, according to at what stage they act --- production, propagation, detection --- and what observables they affect --- energy spectrum, arrival directions, flavor composition, arrival times --- shown as lines connected to the models.  The list of models is representative.}
 \label{fig:classification}
\end{figure}

Given the wide spread of models of new neutrino physics, it is useful to organize them.  Figure \ref{fig:classification} shows our proposed model classification scheme, applied to a representative list of new-physics models.  The scheme classifies a model according to two features: during what stage in the life of the neutrino it acts --- production, propagation, detection --- and what neutrino observables it affects.  A model may act during more than one stage, and may affect more than one observable.  The representative list of models in \figu{classification} shows that many models are able to affect two or three observables, and that most of them act during propagation.

\section{How well can we measure the neutrino observables?}

Statistical and systematic experimental limitations complicate extracting fundamental physics from high-energy cosmic neutrinos.  However, already today, these limitations are surmountable.  In the next decade, larger detectors and improved detection techniques will mitigate them further.

Presently, the main limitation is statistical: after 8 years, IceCube has only detected about 100 contained events, a large fraction of them from neutrinos most likely of cosmic origin.  Several larger neutrino telescopes, currently under construction, will vastly improve the situation: IceCube-Gen2\ \cite{Aartsen:2014njl} --- with 5 times the volume of IceCube --- KM3NeT\ \cite{Adrian-Martinez:2016fdl}, and Baikal-GVD\ \cite{Shoibonov:2019gfj}.  Even larger detectors\ \cite{Olinto:2017xbi, Alvarez-Muniz:2018bhp, RNO:Talk}, in planning, could discover neutrinos with energies 1000 times higher.

Systematic uncertainties in the measurement of the four observables from Section\ \ref{section:landscape} can be significant.  Present neutrino telescopes reconstruct the energy $E$ of a detected event to within 0.1 in $\log_{10}(E/{\rm GeV})$. The arrival direction can be determined  with sub-degree resolution for $\nu_\mu$-initiated tracks, and to within a few degrees for showers initiated mainly by $\nu_e$ and $\nu_\tau$. Arrival time can be measured with  ns precision \ \cite{Aartsen:2013vja}.  Measuring flavor is challenging\ \cite{Mena:2014sja, Palomares-Ruiz:2015mka, Aartsen:2015ivb, Aartsen:2015knd}, since showers initiated by $\nu_e$ and $\nu_\tau$ are similar.  As a result, the IceCube flavor contours in \figu{ternary_plot} are wide.  Progress in improving the measurement of all the observables is ongoing.

Finally, we have entered the multi-messenger era\ \cite{Halzen:2019qkf, Meszaros:2019xej}: the sky is seen in light across many wavelengths, cosmic rays, neutrinos, and gravitational waves.  Their joint study reduces astrophysical unknowns present when extracting fundamental physics from cosmic neutrinos.


\section{Summary: shifting from predictions to tests}

Prior to the discovery of high-energy cosmic neutrinos, there were innovative proposals envisioning their use to study fundamental physics.  However, due to the absence of data, these proposals were mostly limited to being predictions and were weighed down by the large uncertainties of unavoidable educated guesswork.  The IceCube discovery finally materialized the opportunity to put these predictions to test\ \cite{Gaisser:1994yf, Ahlers:2018mkf}.  Today, we may finally switch from primarily making predictions of high-energy new-physics effects to primarily testing predictions against real data. 

\smallskip
\smallskip
\smallskip
\smallskip

{
\footnotesize
{\bf Acknowledgements.} We thank Janet Conrad for useful discussions. 
CA is supported by NSF grant PHY-1801996.
MB is supported by the Villum Fonden project no.~13164.
AK is supported by NSF under grants PLR-1600823 and PHY-1607644 and by the University of Wisconsin Research Council with funds granted by the Wisconsin Alumni Research Foundation.
SPR is supported by a Ram\'on y Cajal contract, by the Spanish MINECO under grants FPA2017-84543-P and SEV-2014-0398, and partially by the Portuguese FCT through the CFTP-FCT Unit 777 (UID/FIS/00777/2019).
JS is supported by the Mar{\'{\i}}a de Maetzu grant MDM-2014-0367 of ICCUB. SPR and JS are also supported by the European Union's Horizon 2020 research and innovation program under the Marie Sk\l odowska-Curie grant agreements No. 690575 and 674896.
ACV is supported by the Arthur B. McDonald Canadian Astroparticle Physics Research Institute. Research at Perimeter Institute is supported by the Government of Canada through the Department of Innovation, Science, and Economic Development, and by the Province of Ontario through the Ministry of Research and Innovation.
}



\providecommand{\href}[2]{#2}\begingroup\raggedright\endgroup

\end{document}